\documentclass[aps,prl,epsfig,showpacs,floats,twocolumn,superscriptaddress,amssymb,amsmath,floatfix]{revtex4}
\usepackage{graphicx}
\begin{document}

\title{Synchronization and Collective Dynamics in A Carpet
of Microfluidic Rotors}
% of Coupled Rotors with Hydrodynamic Interaction}
%

\author{Nariya Uchida}
\email{uchida@cmpt.phys.tohoku.ac.jp}
\affiliation{Department of Physics, Tohoku University, Sendai, 980-8578, Japan}

\author{Ramin Golestanian}
\email{r.golestanian@sheffield.ac.uk}
\affiliation{Department of
Physics and Astronomy, University of Sheffield, Sheffield S3 7RH, UK}

\date{\today}

\begin{abstract}
We study synchronization of an array of rotors on a substrate
that are coupled by hydrodynamic interaction. Each rotor, which is modeled
by an effective rigid body, is driven by an internal torque
and exerts an active force on the surrounding fluid.
The long-ranged nature of the hydrodynamic interaction between the
rotors causes a rich pattern of dynamical behaviors including
phase ordering and self-proliferating spiral waves.
%The model provides a novel example of coupled oscillators with long-range interaction.
Our results suggest strategies for designing controllable microfluidic mixers
using the emergent behavior of hydrodynamically coupled active components.
\end{abstract}


\maketitle

\paragraph{Introduction.}

Microorganisms and the mechanical components of the cell motility
machinery such as cilia and flagella operate in low Reynolds number
conditions where hydrodynamics is dominated by viscous forces
\cite{purcell}. The medium thus induces a long-ranged hydrodynamic
interaction between these active objects, which could lead to
emergent many-body behaviors. Examples of such cooperative dynamical
effects include sperms beating in harmony \cite{Kruse}, metachronal waves in cilia
\cite{metachron,Joanny,Jens}, formation of bound states between rotating
microorganisms \cite{goldstein}, and flocking behavior of red blood
cells moving in a capillary \cite{GompperPNAS}. For a collection of free swimmers,
such as microorganisms \cite{ped-rev}, hydrodynamic interactions have been shown to lead to
instabilities \cite{instability,shelley} that can result in complex dynamical
behaviors \cite{shelley,simulation}. In the context of simple microswimmer models
where hydrodynamic interactions coupled to internal degrees of freedom can be studied with
minimal complexity, %\cite{3SS}
it has been shown that the coupling could result
in complex dynamical behaviors such as oscillatory bound states
between two swimmers \cite{Yeomans2}, and collective many-body
swimming phases \cite{Yeomans3,denis}.

A particularly interesting aspect of such hydrodynamic coupling is
the possibility of synchronization between different objects with
cyclic motions \cite{Joanny,Jens,tom,holger,vilfan,netz,lagomarsino,RaySync,yeomans-new}.
This effect has mostly been studied in simple systems such as two interacting objects
or linear arrays and very little is known about possible
many-body emergent behaviors of a large number of active objects with
hydrodynamic coupling. For example, in a recent experiment \cite{Berg},
Darnton {\em et al.} observed chaotic flow patterns with complex vortices above a
carpet of bacteria with their heads attached to a substrate and their flagella free to
interact with the fluid (see also \cite{Kim-Breuer}).
On the other hand, recent advances from micron-scale magnetically
actuated tails \cite{magnetic} to synthetic molecular rotors \cite{feringa}
now allow fabrication of arrays of active tails that can stir up the fluid.
It is therefore very important to explore the possible complexity of
the phase behavior of such an actively stirred microfluidic system.

Here, we consider a simple generic model of rotors \cite{howard} positioned on
a regular 2D array on a substrate and study their collective dynamics numerically.
We find that the long-ranged hydrodynamic interactions could either enhance or
destroy ordering, depending on the degree of a built-in {\em geometric frustration}
that originates from the interaction of the rotors with the fluid.
More specifically, our model adopts a fully synchronized state when the frustration
is weak, and a randomly disordered state when it is maximally frustrated.
Moreover, the dynamics of the system leads to self-proliferating spiral waves
between the above two limiting behaviors. We also take into account thermal
fluctuations of the rotors and map out the phase diagram of the system
as a function of temperature and the degree of frustration.

\begin{figure}[t]
\includegraphics[width=.9\columnwidth]{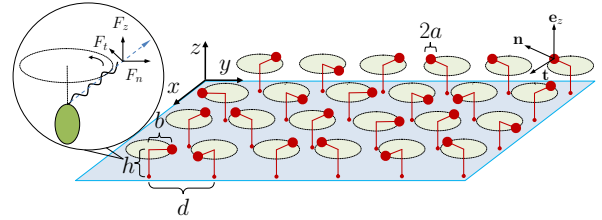}
\caption{(Color online).
Schematic representation of the array of rotors. Inset: an immobilized bacterium
with active flagella as a possible realization of a rotor that can exert both a tangential
drag and an active radial force on the fluid.}
\label{fig:scheme}
\end{figure}

\begin{figure*}[t]
\includegraphics[width=1.99\columnwidth]{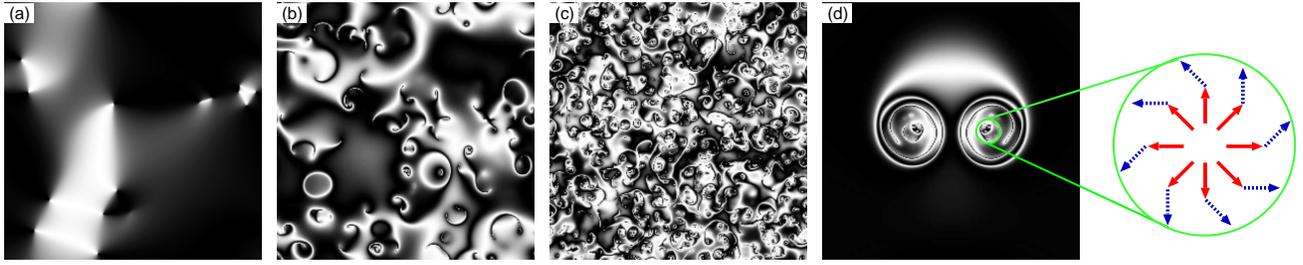}
\caption{(Color online). Snapshots of coarsening defects
(with the greyscale representing $\cos \phi({\bf r})$)
for (a) $\delta=0$ % and (b) $\delta=30^\circ$,
and spiral waves for (b) $\delta=45^\circ$ and (c) $\delta=60^\circ$.
These developed from random initial perturbations.
(d) Spiral waves for $\delta=60^\circ$ evolving from a defect pair,
with a schematic picture of the director field (red, solid arrows)
and the velocity field (blue, dotted arrows) near a $+1$ defect.
}
\label{fig:snapshots}
\end{figure*}

\paragraph{Model and Dynamical Equations.}

We consider an array of rotors that are assumed to be spherical
beads of radius $a$ moving on circular trajectories of radius $b$,
which are positioned on a rectangular lattice of base length $d$ and
at a height $h$ above a substrate (see Fig. \ref{fig:scheme}). The
$i$-th rotor is anchored at ${\bf r}_{i0}$ to the surface of the
substrate, which we take to be the $xy$-plane. The instantaneous
position of the rotating bead is ${\bf r}_i = {\bf r}_{i0} + b {\bf n}_i + h {\bf e}_z$,
where the unit vector ${\bf n}_i(t)=(\cos\phi_i(t), \sin\phi_i(t), 0)$
gives the orientation of the arm of the rotor. Because of the constraint
that the bead is only allowed to move on the circular orbit of
radius $b$, the velocity of the rotor can be written as
${\bf v}_i = b\frac{d{\bf n}_i}{dt} = b \frac{d \phi_i}{d t} {\bf t}_i$,
where ${\bf t}_i= {\bf e}_z \times {\bf n}_i = (-\sin\phi_i, \cos\phi_i, 0)$
is the unit vector tangent to the trajectory.

We assume that the structure of the rotor is such that it drags the
fluid with it as it moves (tangentially) along the circular
trajectory, while it can also pump the fluid radially due to some
internal degrees of freedom. The inset of Fig. \ref{fig:scheme}
shows a possible realization of such a system in the case of
bacteria whose heads are fixed on the substrate. In this example,
the spinning rotation of the flagella would produce the pumping
effect, while the precession of the axis of the flagella about the
anchoring point would correspond to the tangential motion of the
bead. Therefore, in our simplified model each rotor exerts a force,
which can be decomposed into the radial, tangential, and vertical
components as
${\bf F}_{i} = F_n {\bf n}_i + F_t {\bf t}_i + F_z {\bf e}_z$.
The velocity field of the fluid created by the rotors is given by
\begin{math}
{\bf v}({\bf r}) = \sum_i {\bf G}({\bf r} - {\bf r}_{i}) \cdot {\bf F}_{i}
\end{math}
where ${\bf G}({\bf r})$ is the Blake-Oseen tensor~\cite{Oseen},
which describes the hydrodynamic interaction near a flat surface
with the non-slip boundary condition.
Assuming that the arm length $b$ and the height $h$
are much smaller than the characteristic distance $d$ between
the rotors, we can use
the $O(h^2/d^2)$ approximation~\cite{Joanny},
\begin{math}
G_{\alpha\beta}({\bf r})=\frac{3h^2}{2\pi\eta} \frac{r_\alpha r_\beta}{|{\bf r}|^5}
\end{math}
for $\alpha,\beta = x,y,$ and
\begin{math}
G_{\alpha z}({\bf r})=G_{z\alpha}({\bf r})=G_{zz}({\bf r})=0,
\end{math}
for $\alpha = x,y$. Note that the $z$-component of the force is
not coupled to flow and that the fluid velocity is lying in the $xy$-plane.

To obtain the flow velocity at the position of the rotors, we need to subtract
the self-interaction, which involves the Stokes drag coefficient $\zeta = 6\pi \eta a$.
This yields
\begin{equation}
\frac{d \phi_i}{d t}=\omega_t+ \frac{3 \gamma}{2\pi} \sum_{j\neq i}
\frac{{\bf t}_{i} \cdot {\bf r}_{ij} \;{\bf r}_{ij}
\cdot \left(\omega_n {\bf n}_j + \omega_t {\bf t}_j \right)
}{|{\bf r}_{i j}|^5},
\label{dotphi}
\end{equation}
where ${\bf r}_{ij}={\bf r}_i-{\bf r}_j$, $\omega_{t,n}=F_{t,n}/(\zeta b)$
are the reduced forces, and
\begin{math}
\gamma =\zeta h^2/\eta =  6\pi a h^2
\end{math}
is the hydrodynamic coupling constant.
When the interaction is weak, we can simplify the phase equation
(\ref{dotphi}) following a standard prescription~\cite{Kuramoto}.
To this end, we rewrite it 
in terms of the slow variable $\Phi_i = \phi_i-\omega_t t$,
and then integrate it over a cycle under the approximation that
$\Phi_i$ in the interaction term is constant over the period $2\pi/\omega_t$.
We obtain
\begin{eqnarray}
\frac{d \Phi_i}{d t} = - \frac{3 \gamma\omega}{4\pi} \sum_{j\neq i}
\frac{1}{|{\bf r}_{i j}|^3}
\sin(\Phi_i - \Phi_j - \delta),
\label{dotPhi}
\end{eqnarray}
where $\delta=\tan^{-1}(\omega_t/\omega_n)$ and
$\omega=\sqrt{\omega^2_t + \omega^2_n}$.
This equation is correct to $O(\gamma /(d^3\sin \delta))$~\cite{Kuramoto,hsync-paper4}.
In this form, the hydrodynamic coupling becomes isotropic and
our system resembles existing models of non-locally coupled
oscillators with phase delay~\cite{Shima,Kim}.

\paragraph{Simulation Method.}

\begin{figure*}[t]
\includegraphics[width=1.99\columnwidth]{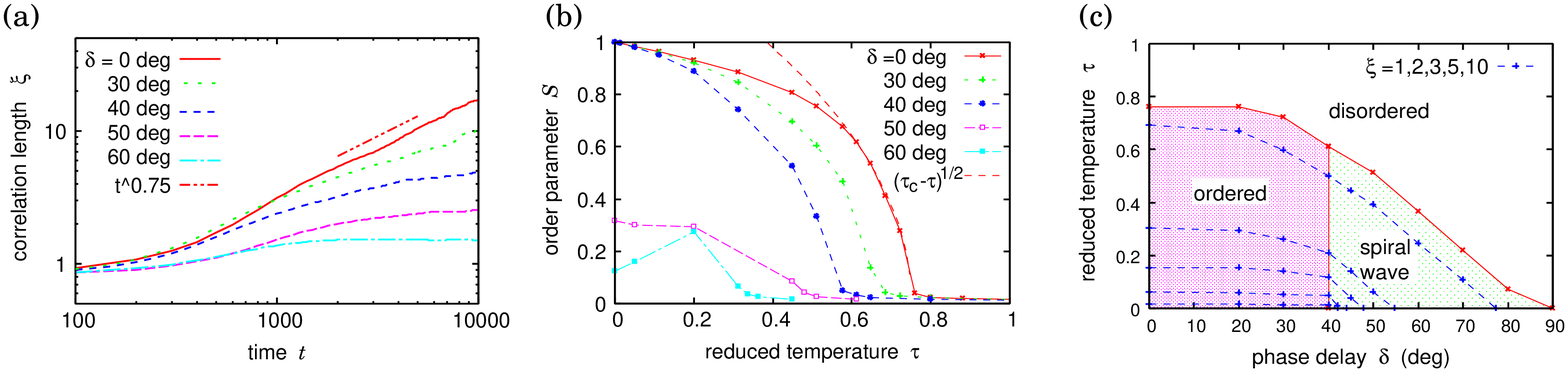}
\caption{(Color online). (a) Correlation length $\xi$ as a function of time and for different values
of $\delta$. For $\delta=0^\circ$ it is well fitted by $\xi \propto t^{0.75}$.
For $\delta \le \delta_c=40^\circ$ the correlation length diverges after
$t=10000$ (not shown), while for $\delta> \delta_c$ it remains finite.
(b) Equilibrium order parameter $S$ as a function of temperature $\tau$
and for different values of $\delta$.
(c) Phase diagram and contour of the correlation length $\xi$.
The correlation length is smaller for larger $\delta$
and higher temperature.
}
\label{fig:comb}
\end{figure*}

The model is implemented on a $L\times L$ square lattice
with the grid size $d$, and the phase equation Eq. (\ref{dotphi})
is solved by Euler method with the time step $\Delta t$.
The system size used is $L=256$ for most of the results below,
while we have also used $L=128$ for obtaining some of the statistical data.

We have imposed periodic boundary condition and computed
the velocity field at every time step by Fourier transformation.
For $\delta>0$, we also solved the reduced phase equation Eq. (\ref{dotPhi}),
to compare with the solution of Eq. (\ref{dotphi}), and found a very good agreement.
We have also incorporated thermal fluctuations, by adding an uncorrelated
Gaussian noise $\Omega_{i}(t)$ to the RHS of Eq. (\ref{dotphi}).
The noise is assumed to have zero mean and its fluctuations are
controlled by the rotational diffusion constant of the bead
$D_r = k_{\rm B} T/(\zeta b^2)$ as
\begin{math}
\left\langle \Omega_{i}(t) \Omega_{j}(t') \right \rangle
=2 D_r \delta_{ij} \delta(t-t').
\end{math}
We define
% the dimensionless coupling constant $g = \gamma/d^3$ and
the reduced (effective) temperature
\begin{math}\displaystyle
\tau = \frac{D_r d^3}{\gamma \omega}
= \frac{k_BT d^3}{36 \pi^2 \eta a^2 b^2 h^2 \omega}.
%\label{temperature}
\end{math}
For typical values of $a \sim b \sim h \sim 1 \;\mu$m,
$d \sim 10 \;\mu$m, $\omega \sim 10^2$ Hz, with
$\eta = 1 \times 10^{-3}$ Pa s and $k_B T = 4 \times 10^{-21}$ J,
we have
% $\zeta \sim 10^{-8}$ kg/s, $D_r \sim 10^{-1}$ Hz, $g \sim 10^{-2}$ and
$\tau \sim 10^{-1}$.
Note that the sharp dependence of $\tau$ on $a$, $b$, $h$, and $d$ makes it easy to
control the reduced temperature by changing the size or density of rotors.
In our simulations, we have used the parameter values
$\gamma=0.1$ and $\omega=0.1$, with $d=1$
and $\Delta t=0.1$.
First we turn off the thermal noise ($\tau=0$) and
vary the force angle $\delta$ to study the pattern dynamics.

\paragraph{Pumping-driven Rotors.}

When $\delta=0$, each rotor pumps the fluid radially, and
is driven by the fluid flow generated by the other rotors.
In this case, the initial random perturbations
develop into topological defects
(singularities of the phase field $\phi({\bf r}$)
of winding numbers $\pm 1$,
which coarsen by collision of $+1$- and $-1$- defects and
finally disappear to establish global synchronization.
Figure \ref{fig:snapshots}(a) shows a snapshot of the coarsening defects
at $t=10000$. To characterize the phase ordering dynamics,
we define the correlation length
\begin{math}
\xi = {\langle (\nabla \phi({\bf r}))^2 \rangle}_{\bf r}^{-1/2},
\end{math}
where ${\nabla} \phi$ is the spatial gradient
and $\langle ...\rangle_{\bf r}$ means spatial average.
As shown in Fig. \ref{fig:comb}(a), we find that $\xi$ as a function of time
is well fitted by the power law $\xi \propto t^\nu$, with $\nu=0.75$.
% The coarsening exponent is larger than that of the diffusive XY model~\cite{Blundell-Bray},
% $\nu=0.37$ (numerical) and $\nu=1/2$ (scaling).
% In fact,
The scaling of the hydrodynamic interaction
${\bf G}({\bf r}) \sim |{\bf r}|^{-3}$ and Eq. (\ref{dotphi})
suggest that the characteristic timescale of a pattern
is proportional to its size, which would mean $\nu=1$.
The difference between the numerical and scaling exponents
suggests violation of dynamic scaling,
% due to logarithmic correction of 2D topological defects
% as seen in the XY model
% which is
which is characteristic to coarsening of point defects
in two dimensions~\cite{Blundell-Bray}.

\paragraph{Torque-driven Rotors.}

When $\delta= 90^\circ$, each rotor is driven by
an active torque and exerts a force tangential to its orbit.
In this case, we find that the system reaches a disordered state
in which spatial correlation is almost completely lost.
% The orientational correlation function
% $C_\phi({\bf r}) =  \langle
% \cos(\phi({\bf r}+{\bf r'}) \cos(\phi({\bf r'})))
% \rangle_{\bf r'}$
% has a $\delta$-function like peak at ${\bf r}=0$ with
% a very low long tail that scales with $r^{-3/2}$ (not shown).
The absence of orientational correlation can be understood as follows.
The average flow created by a rotor is perpendicular to its arm,
and a neighboring rotor tends to align with the flow.
Thus the two rotors tend to be perpendicular to each other
on average. However, it is not possible that every pair of rotors
have their arms vertically crossed (geometric frustration),
and hence the system evolves towards randomly oriented states.

\paragraph{Rotors driven by Pumping and Torque.}

In the general case, the rotor is driven by an active torque
while it pumps the fluid radially, and the total force
exerted on the fluid has an angle $0^\circ < \delta < 90^\circ$
with respect to the arm of the rotor.
We varied the parameter $\delta$ and found
two types of dynamical behavior.
For $0^\circ < \delta \le 40^\circ$,
the globally synchronized state is still obtained as the final
state. %[see Fig. \ref{fig:snapshots}(b)].
However, for $40^\circ < \delta < 90^\circ$, we find that the dynamical steady
state of the system involves self-proliferating spiral waves as shown in Fig.
\ref{fig:snapshots}(b) and (c). Moreover, we find that the correlation length
$\xi(t)$ converges to a finite value as $t\to \infty$ as shown in Fig. \ref{fig:comb}(a),
and that the equilibrium correlation length decreases as $\delta$ is increased.

The flow pattern is locally correlated with the rotor's director
${\bf n}({\bf r})= (\cos \phi({\bf r)}, \sin \phi({\bf r}))$.
The surface flow velocity ${\bf v}({\bf r})$ makes the angle $\delta$
with ${\bf n}({\bf r})$ except at the core of the defect (see the movies~\cite{movies}
for comparison of the two fields).
This observation leads us to an intuitive interpretation of the spiral waves.
In the vicinity of a $+1$ defect from which the director emanates radially,
the rotors create an outgoing flow that has an anti-clockwise slant 
with respect to
the radial direction, and hence form an anti-clockwise spiral;
see the inset of Fig.\ref{fig:snapshots}(d).
The spiral is tighter for a larger force angle $\delta$.
We confirmed this scenario by choosing a defect pair
as the initial configuration and following its evolution;
see Fig.\ref{fig:snapshots}(d) and the corresponding movie \cite{movies}.
Initially, clockwise and anti-clockwise spirals are
formed around $-1$- and $+1$-defects, respectively.
Then the director is randomized on the thinning spiral arm,
which collapses and proliferates a cascade of new defects.

% For $\delta>\delta_c=40^\circ$, the defect pair
% collides and annihilates for $\delta \le \delta_c = 40^\circ$, while
% it
% develops into self-proliferating spiral waves for $\delta > \delta_c$.
% For the latter case,

\paragraph{Thermal Fluctuations.}

We then introduce the thermal torque and study the phase behavior of
the system. In Fig. \ref{fig:comb}(b), we plot the equilibrium order parameter
$S = |\langle \cos\phi \rangle|$ as a function of the effective temperature.
For $\delta=0$, we find a critical temperature $\tau_c$
at which $S$ vanishes as $S \propto \sqrt{\tau_c-\tau}$ to a good approximation \cite{note1}.
We find the critical temperature for this phase transition as $\tau_c =0.76$,
which is about $30\%$ smaller than the mean-field value $\tau_c = 1.08$
by Guirao and Joanny~\cite{Joanny}.

As we increase the phase delay $\delta$ up to $\delta_c=40^\circ$,
the critical temperature is lowered down to $\tau_c=0.61$.
For $\delta>\delta_c$, the order parameter $S$ is less than $1$
even at $\tau=0$ and is smaller for a larger system size $L$,
suggesting that $S=0$ in the thermodynamic limit.
However, the existence of local order (spiral waves) is reflected
in the finite-$L$ data, using which we can define the transition
temperature $\tau_c$ in the same way as mentioned before \cite{note1}.
The resulting phase diagram is shown in Fig. \ref{fig:comb}(c).
We distinguish three regions: (O) {\em ordered}, which is a distinct thermodynamic phase characterized by global synchronization, and (S) {\em spiral waves} and (D) {\em disordered}, which are inherently the same phase but with different local ordering and dynamical structure.
Also shown in Fig. \ref{fig:comb}(c) are the contours of the correlation length $\xi$.
While the O-D and O-S transitions are sharp, the S-D transition
is a crossover characterized by gradual decrease of $\xi$.
We also note that the frustrated state for $\delta=90^\circ$, $\tau=0$
and the thermally-disordered state for $\delta=0^\circ$, $\tau>\tau_c$ are
qualitatively different, though they are not distinguished
in the phase diagram.

\paragraph{Discussion.}

The case of no active torque ($\delta=0$) could be regarded as
a simplified and idealized model of bacterial carpets~\cite{Berg,Kim-Breuer}.
Our model reproduced the enhancement of orientational ordering,
while it predicts global ordering and not the finite-size correlation
as observed in the experiments. The experimental patterns might be explained
by some kind of frozen disorder in the flagellar configuration,
which can be readily incorporated in our model~\cite{hsync-paper2,hsync-paper4}.
%We have mapped out the phase diagram for the collective behavior of an array of rotors
%that are coupled by hydrodynamic interactions as a function of
%the force angle $\delta$ and the reduced temperature $\tau$.
% We note that for the special
% For the case of $\delta=90^\circ$, the tangential force of
% each rotor generates an average flow perpendicular to the rotor's arm and
% thus works against local ordering. This
The case of $\delta=90^\circ$ is realized by rigid spheres without pumping.
It is related to a recently studied model of two rigid spheres making
tilted elliptic orbits \cite{vilfan}, which show both
in-phase and anti-phase synchronization.
% depending on the mutual orientation of the two orbits.
Our results suggest that the interaction between many of such rotors
is frustrated and the system does not attain full synchronization.
%This is also supported
%by linear stability analysis of Eq. (\ref{dotPhi}), which implies that for any
%$\delta < \pi/2$, and not for $\delta=\pi/2$, the globally synchronized state is
%linearly stable. Our numerical results, however, show that for $\delta > \delta_c$
%the proliferation of the spiral defects (waves) and their frustrated dynamics does not allow
%the system to reach the globally synchronized state.
Spiral waves for finite phase delay $\delta$ have been observed
in previous models of 2D coupled oscillators~\cite{Kim,Shima,Sakaguchi}.
However, in our case, the pattern is intrinsically turbulent and self-proliferating,
in contrast to the case of finite-range coupling, for which the phase is spatially smooth except
near the defect core~\cite{Sakaguchi,Kim,Shima}.

In conclusion, we have introduced a generic model of microfluidic rotors
that shows a variety of dynamical patterns including global synchronization,
fully disordered states, and self-proliferating spiral waves.
The patterns are sensitively controlled by the angle of active force $\delta$
(the degree of frustration) and the temperature $\tau$.
%Our model lays a basis for understanding the complex behaviors
%observed in bacterial carpets~\cite{Berg}.
% The turbulent regime has similarities to the complex fluid
% patterns observed recently in bacterial carpets
% The relevance of thermal fluctuation is measured
% by our reduced temperature, %[Eq. (\ref{temperature})]
% which suggests that the synchronized-disorder transition is easily achievable
% in practice by controlling the density of the microfluidic rotors.
Our results suggest that arrays of active microfluidic components could be
designed to induce a rich variety of dynamical behaviors in the vicinal
fluid, and could be used to make switchable microfluidic mixers.

\acknowledgments

NU thanks the hospitality at University of Sheffield
where this work was initiated, and financial support
from Grant-in-Aid for Scientific Research 
from MEXT. RG acknowledges financial support from the EPSRC.


\begin{thebibliography}{99}
\bibitem{note1}
In practice, $S$ remains finite even for $\tau > \tau_c$ due to
the finite-size effect; we defined $\tau_c$ by $S(\tau_c) = 2 S_0$ where
$S_0$ is the theoretical value of $S$ for a non-interacting
($\gamma=0$) system.
\end{thebibliography}
\end{document}